\documentstyle[12pt,aasms4]{article}
\lefthead{Ikuta  et al.}
\righthead{Why CO in high-$z$ quasar ?}
\accepted{PASJ}
\begin{document}

\title{WHY CAN WE DETECT QUASAR BR1202$-$0725 IN CO ?}
\author{Chisato Ikuta, Nobuo Arimoto, \& Yoshiaki Sofue}

\affil{Institute of Astronomy, University of Tokyo, 2-21-1 Osawa,
Mitaka, Tokyo 181, Japan; ikuta@mtk.ioa.s.u-tokyo.ac.jp,
arimoto@mtk.ioa.s.u-tokyo.ac.jp, sofue@mtk.ioa.s.u-tokyo.ac.jp}

\author{Yoshiaki Taniguchi}

\affil{Astronomical Institute, Tohoku University, Aoba, Sendai 980-77, 
Japan; tani@astroa.astr.tohoku.ac.jp}

\begin{abstract}

We present CO luminosity evolution of both elliptical and spiral galaxies
based on a galactic wind model and 
a bulge-disk model, respectively.
We have found that the CO luminosity peaks around the epoch of galactic wind
caused by collective supernovae $\sim$ 0.85 Gyr after the birth of
the elliptical with $M = 2 \cdot 10^{12} M_\odot$ while
$\sim$ 0.36 Gyr after the birth of
the bulge with $M = 2 \cdot 10^{11} M_\odot$.
After these epochs, the CO luminosity
decreases abruptly because the majority
of molecular gas was expelled from the galaxy system as the wind.
Taking account of typical masses of elliptical galaxies and bulges
of spiral galaxies, we suggest that CO emission can be hardly
detected from galaxies with redshift $z \sim 1 - 4$ unless some
amplification either by galaxy mergers and/or 
by gravitational lensing is working.
Therefore, our study explains reasonably why CO emission was detected
from the high-redshift quasar
BR 1202$-$0725 at $z =4.7$ 
while not detected from the powerful radio galaxies with $1 < z < 4$. 

\end{abstract}

\keywords{cosmology: observations {\em -} galaxies: evolution {\em -} 
galaxies: formation {\em -} galaxies - nuclei of {\em -} Radio lines:
CO {\em -} Radio sources: AGNs}

\section{INTRODUCTION}

The formation and evolution of galaxies is one of the
fundamental problems in astrophysics.
The recent deep imaging of very faint galaxies
made with {\it Hubble Space Telescope} (Williams et al. 1996)
and the detection of CO emission from a high-$z$ quasar BR 1202$-$0725
at $z$ = 4.69 (Ohta et al. 1996; Omont et al. 1996) have encouraged 
us to study the problem mentioned above.
Since the galaxies 
should form from gaseous system, it is important to
investigate the major epoch of star formation in the gas system and 
to study how stars have been made during the course of galaxy evolution.
When we study evolution of galaxies, we usually use stellar lights
as the tracer of evolution (cf. Tinsley 1980; Arimoto \& Yoshii 1986, 1987;
Bruzual \& Charlot 1993). However, much data of interstellar medium (ISM)
of galaxies from X-ray emitting hot gas through warm HI gas to cold molecular 
gas and dust have been accumulated for these decades 
(cf. Wiklind \& Henkel 1989; Lees et al. 1991;
Fabbiano, Kim, \& Trinchier 1992; Kim, Fabbiano, \& Trinchier 1992; 
Wang, Kenney, \& Ishizuki 1992).
Therefore, the time is ripe to begin the study of evolution of ISM of galaxies
from the epoch of galaxy formation to the present day for both elliptical
and disk galaxies.

In this {\it Paper}, appreciating the recent detection of CO emission
from the high-$z$ quasar BR 1202$-$0725 (Ohta et al. 1996; Omont et al. 1996),
we discuss the evolution of molecular gas content in galaxies. 
Since active galactic nuclei (AGN) are associated with their host galaxies,
the CO luminosity of AGN depends
on the gaseous content of their host galaxies.
Therefore, any observations of molecular-line emission from high-$z$ objects
are very useful in studying evolution of molecular gas content
in galaxies
\footnote{Besides the CO detection from BR 1202$-$0725, there are 
two more successful detections  of CO from high-$z$ objects;
1) the hyperluminous infrared galaxy
IRAS F10214+4725 at $z$ = 2.286
(Brown \& Vanden Bout 1991; Solomon, Downes, \& Radford 1992; 
Tsuboi \& Nakai 1992; Radford et al. 1996), and 2) the Cloverleaf quasar  
H1413+135 at $z$ 2.556 (Barvanis et al. 1994).
Since, however, these two sources are gravitationally amplified ones
(Elston, Thompson, \& Hill 1994;
Soifer et al. 1995; Trentham 1995; Graham \& Liu 1995;
Broadhurst \& Leh\'ar 1995; Serjeant et al. 1995;
Close et al. 1995), we do not use these data in this study
because there may be uncertainty in the
amplification factor.}.
In spite of the successful CO detection from BR 1202$-$0725, 
Evans et al. (1996) reported the negative CO detection from 11 high-$z$
($1 < z < 4$) powerful radio galaxies (PRGs)
and gave the upper limits of order 
$M_{\rm H_2} \sim 10^{11} M_{\odot}$ 
(hereafter $H_0$ = 50 km s$^{-1}$ Mpc$^{-1}$
and $q_0 = 0.5$), being comparable to or larger than that of BR 1202$-$0725
(Ohta et al. 1996; Omont et al. 1996).

Here a question arises as why CO emission was detected from the high-$z$
quasar at $z = 4.69$ while not detected from the high-$z$ ($1 < z < 4$)
PRGs. There may be two alternative 
answers: 1) The host galaxies are different
between quasars and PRGs in terms of molecular gas contents. Or, 2) although 
the host galaxies are basically similar between quasars and PRGs, their
evolutionary stages are different and thus the molecular gas contents are
systematically different between the two classes.
Provided that the current unified model for quasars and radio galaxies
(Barthel 1989) is also applicable to high-$z$ populations, 
it is unlikely that their host galaxies are significantly different. 
Since it is usually considered that luminous AGNs like quasars
and PRGs are associated with
either massive ellipticals or bulges of disk galaxies 
as well as merger nuclei, 
the evolution of
ISM would be rapidly proceeded during the era of spheroidal
component formation. Therefore, we investigate the latter possibility 
(i.e., evolutionary effect) and discuss some implications on the evolution
of ISM in galaxies.

\section {MODELS}

Assuming that the luminous AGNs 
are harbored in giant elliptical galaxies and/or in bulges of spiral galaxies, 
we investigate the evolution of CO luminosity 
based on a galactic wind model for 
elliptical galaxies proposed by Arimoto \& Yoshii (1987) and 
a bulge-disk model for spiral galaxies by Arimoto \& Jablonka (1991). 
The so-called {\it infall} model of galaxy chemical evolution is adopted for
both spheroidals and disks and time variations of gas mass and gas 
metallicity, in particular $\log ({\rm O/H})$, are calculated numerically 
by integrating usual differential equations for chemical evolution
without 
introducing the instantaneous recycling approximation for stellar lifetime. 
Model parameters, such as star formation
rate (SFR) $k$, a slope of initial mass function (IMF) $x$, and 
gas accretion rate (ACR) $a$, are
taken from Arimoto \& Yoshii (1987) and Arimoto \& Jablonka (1991). 
The lower and upper stellar mass limits are set 
to be $m_{\ell}=0.1$ M$_{\odot}$ and $m_u=60$ M$_{\odot}$, respectively.

According to Jablonka, Martin, \& Arimoto (1996), 
who found that the $Mg_2-{\rm log} \sigma$ relation of bulges are 
exactly identical to that of elliptical galaxies, 
we consider that bulges are
small ellipticals of equivalent luminosity
and that both spheroidal systems share the similar history of star formation. 
Thus, for ellipticals and bulges,
we assume that the remaining gas is expelled completely
after the onset of galactic wind, which takes place once the thermal 
energy released from supernovae exceeds the binding energy of the gas.
The wind times, $t_{\rm gw}=0.85$ Gyr for giant ellipticals 
(M$_{\rm init}=2 \cdot 10^{12}$ M$_{\odot}$) and $t_{\rm gw}=0.36$
Gyr for bulges (M$_{\rm init}=2 \cdot 10^{11}$ M$_{\odot}$),
are taken from Arimoto \& Yoshii (1987).

For spiral galaxies, assuming that the bulge and disk evolve independently,
we construct a model by combining the bulge and disk models with
M$_{\rm init}=2 \cdot 10^{11}$ M$_{\odot}$ and $10^{11}$ M$_{\odot}$,
respectively. This model gives $M_{\rm v}=-20.97$ mag for the bulge and
$M_{\rm v}=-20.87$ mag for the disk at the age of 15 Gyr old 
(Arimoto \& Jablonka 1991).
The bulge-to-disk light ratio in V-band is $L_b/L_d \simeq 1$, 
nearly twice of typical values for early type spirals (Simien \& 
de Vaucouleurs 1986).

The $L_{\rm CO}$\footnote{The CO luminosity $L_{\rm CO}$ refers to that of
CO($J$=1-0). 
Note that the $L_{\rm CO}$ of high-$z$ galaxies are measured by using
much higher transitions such as $J$=3-2, 4-3, and so on. However,
it is known that that local
CO-rich galactic nuclei and starburst nuclei have
$L_{\rm CO}$($J$=3-2)$/L_{\rm CO}$($J$=1-0) $\simeq 1$ (Devereux et al. 1994;
Israel \& van der Werf 1996). Therefore high-$z$
analogs may have the similar properties. In fact, two high-$z$ objects
IRAS F10214+4724 and H1413+117 have
$L_{\rm CO}$($J$=4-3)$/L_{\rm CO}$($J$=3-2) $\simeq 1$ and 
$L_{\rm CO}$($J$=6-5)$/L_{\rm CO}$($J$=3-2) $\simeq 0.6-1$ (see Table 1 of
Israel \& van der Werf). 
Thus, the uncertainty due to use of higher transition data
may be 50 percent at most, when we compare model $L_{\rm CO}$($J$=1-0) and
the observed $L_{\rm CO}$ at higher transitions.}
of a model galaxy 
can be calculated from molecular hydrogen mass 
by using the empirical CO--to--H$_2$ conversion factor ($X^*$). 
Arimoto, Sofue \& Tsujimoto (1995) 
showed that $X^*$ strongly depends on the gas metallicity 
and derived the following relationship valid for nearby
spirals and irregular galaxies:
\begin{equation}
{\rm log} X^*=-1.0 (12 + {\rm log} {\rm O/H}) + 9.30,
\end{equation}
where $X^* \times 10^{20}$  H$_2$/ K km s$^{-1}$
=$N_{\rm H_2}/I_{\rm CO}$ and
O/H is the oxygen abundance of HII regions.
We introduce a fractional mass of hydrogen molecule to that of atomic
hydrogen, $f_{\rm mol} \equiv M_{\rm H_2}/M_{\rm HI}$, 
and write the CO luminosity in K km s$^{-1}$ pc$^2$
as follows:
\begin{eqnarray}
1.6 \cdot X^* \cdot L_{{\rm CO}}=
M_{{\rm H_2}} = M_{{\rm HI}} \cdot f_{{\rm mol}} 
\end{eqnarray}
where $M_{{\rm HI}}$ and $M_{{\rm H_2}}$ are in $M_\odot$. 

Chemical evolution model gives $M_{\rm HI}$ 
and O/H as a function of 
time and the CO luminosity evolution can be traced with a help of 
Eq.(2) provided that $f_{{\rm mol}}$ is known {\it a priori}.
We assume time invariant $f_{{\rm mol}}$ throughout
the course of galaxy evolution. In principle, $f_{{\rm mol}}$ itself
should evolve as well, since the hydrogen molecule is newly produced
on the surface of dust ejected from evolving stars and/or formed
in expanding shells of supernovae remnants
while at the same time a part of molecules are dissociated by
UV photons emitted from young hot stars.
The mass of dust and the number of UV photons should also
evolve as a result of galactic chemical evolution 
(Honma, Sofue, \& Arimoto 1995). 
Detailed evolution of $f_{{\rm mol}}$ will be shown in our subsequent
paper (Ikuta et al. 1997),
instead in this {\it Paper} we assume $f_{{\rm mol}}=0.2$. Recent studies
of nearby ellipticals suggest $f_{{\rm mol}} \simeq 0.05 - 0.5$ 
(Wiklind \& Rydbeck
1986; Sage \& Wrobel 1989; Lees et al. 1991; 
Eckart, Cameron, \& Genzel 1991).
The contribution of helium to the gas mass is entirely ignored for 
simplicity, but our conclusions change little even if the evolution 
of helium gas is precisely taken into account.

The formation epoch of galaxies is assumed to be 
$z_f=10$. Although the choice of $z_f$ is rather arbitrary, 
$z_f \ge 5$ has some supports from
recent studies on the metallicity 
of broad emission-line regions of high-$z$ quasars 
(Hamann \& Ferland 1992, 1993; Kawara et al. 1996; Taniguchi et al. 1997).

\section {RESULTS}
\subsection {CO in ellipticals}

Figure 1 shows the result for elliptical galaxies.
The thick solid line represents the galactic wind model, and the dotted
line a model with continuous star formation
(the wind is suppressed even after the wind criterion is satisfied).
The dashed line shows a case for a wind model, but the
gas ejected from evolving giants after the wind
is bound and accumulated in the galaxy to form neutral gas
(bound-wind model; Arimoto 1989).
The CO luminosity, $L_{{\rm CO}}$, of elliptical galaxies increases
prominently soon after their birth, and attains the maximum at
an epoch of about 0.85 Gyr since the birth, or at $z\simeq 4$.
Then, it suddenly decreases when the galactic wind has expelled 
the ISM from the galaxy.

The extremely luminous phase in CO observed for the high $z$
quasar BR 1202$-$0725 
(Ohta et al. 1996; Omont et al. 1996) can be well explained,
if it is in the star forming phase of the whole elliptical system.
Moreover, the non-detection of the smaller redshift galaxies
as observed by Evans et al. (1996) and van Ojik et al. (1997)
 is also naturally understood by
the present model: It is  because of the fact that elliptical galaxies
at $z<4$ contains little ISM.
 
In the figure, we also superpose CO observational data 
for lower redshift elliptical galaxies
(Wiklind \& Rydbeck 1986; Sage \& Wrobel 1989; Wiklind \& Henkel 1989;
Eckart et al. 1991;  Lees et al. 1991; Sage \& Galletta 1993;
Sofue \& Wakamatsu 1993; Wiklind, Combes, \& Henkel 1995).
The theoretical curve for the bound-wind model 
is clearly inconsistent with the observations for galaxies
at $z<0.1$.
This suggests that the gas has been expelled continuously
after the galactic wind ($z$ $\simeq$ 4) and has not been
bound to the system.  
This, in turn, is consistent with the idea that the
intracluster hot gas with high metallicity, as observed in
X-rays, may have been supplied by the winds from early type galaxies
(Ishimaru \& Arimoto 1997).
Although it is not clarified how the gas has been
expelled out of the galaxies, without being bound
to the system, 
recent studies suggest that it is 
probably due to the energy supply from 
either the type Ia supernovae (Renzini et al. 1994) or 
the intermittent AGN activities (Ciotti \& Ostriker 1997).  

\subsection {CO in spirals}

Figure 2 shows the result for a spiral galaxy, where 
the initial masses of bulge
and disk are taken to be 
$2 \cdot 10^{11}M_\odot$ and $10^{11}M_\odot$, respectively.
The CO luminosity of the bulge evolves in almost the same fashion
as an elliptical galaxy as above:  $L_{{\rm CO}}$
increases rapidly after the birth, attains the maximum within
0.36 Gyr, and, then, suddenly decreases because of the
strong wind from the star-forming bulge.
The CO luminosity of the thus-calculated forming bulge
seems insufficient to be detected as the observed luminosity 
of BR 1202$-$0725, unless the bulge is much heavier 
than $2 \cdot 10^{11}M_\odot$.
Moreover, we emphasize that the duration of this bright phase 
in $L_{{\rm CO}}$ is shorter than that obtained for
ellipticals by a factor of two, and therefore, the probability to detect
such CO-bright phase for a bulge would be much smaller than that
for elliptical galaxies.

On the other hand, formation of the gaseous disk due to gas infall
and star formation then proceeds mildly, and, therefore,
the metal pollution of ISM in the disk is slower, which
results in a slower increase of the CO luminosity.
As a consequence, the CO luminosity increases gradually and
monotonically until today.
Also, the less-luminous phase due to the disk, 
following the wind phase of the bulge,
is in agreement with the upper-limit observations of
Evans et al. (1996) and van Ojik et al. (1997).

We also plot CO observations for more other nearby spiral
galaxies, as plotted by filled circles (Braine et al. 1993).
The evolution of the CO luminosity of these galaxies can be
traced back by adjusting the present-day luminosity of the
calculated track. The most luminous nearby spirals in CO is NGC 4565
($L_{\rm CO} \simeq 6 \cdot 10^9$ K km s$^{-1}$ pc$^2$).
It is interesting to mention that, if the model is normalized to this 
galaxy, the peak CO luminosity corresponding to the forming bulge phase 
can be still sufficient to explain the luminosity of BR 1202$-$0725.

\section {DISCUSSION}

The present study has shown that the current radio
telescope facilities are capable of detecting CO emission
from high-redshift galaxies which experience their
initial starbursts if the following two 
conditions are satisfied; 1) the masses of systems should
exceed  $\sim 2 \cdot 10^{12} M_\odot$, and 2) their evolutionary phases
should be prior to the galactic wind.
Therefore, the detectability of CO emission from high-$z$ galaxies 
is severely limited by the above two conditions.
Our study suggests that CO emission can be hardly
detected from galaxies with redshift $z \sim 1 - 4$ without an
amplification either by galaxy mergers and/or by gravitational lensing.
This prescription is consistent with the observations;
CO emission was detected
from the high-redshift quasar
BR 1202$-$0725 at $z =4.7$ (Ohta et al. 1996; Omont et al. 1996)
while not detected from the radio galaxies with $1 < z < 4$
(Evans et al. 1996 and van Ojik et al. 1997) 
and quasars with redshift $\sim$ 2 
(Takahara et al. 1984).
Further, the two convincing detections of CO emission from the high-$z$
objects at $z \sim 2$, IRAS F10214+4724 
(cf. Radford et al. 1996) and the Cloverleaf
quasar H1413+135 (Barvanis et al. 1994), are actually 
gravitationally amplified sources. 

The striking non-detection of high-$z$ galaxies in CO at $1 < z < 4$
implies that
most elliptical galaxies and bulges of spiral galaxies were 
formed before $z \sim 4$, or high-$z$ galaxies with $1 < z < 4$ observed
in the optical and infrared studies may be galaxies after the epoch of 
galactic wind.  
This implication is consistent with the formation epoch ($z > 4$)
of high-$z$ quasars studied by chemical properties of the broad emission-line
regions  (Hamann \& Ferland 1992, 1993; Hill, Thompson, \& Elston 1993;
Elston et al. 1994; Kawara et al. 1996; Taniguchi et al. 1997).
Therefore it is strongly suggested that most host galaxies of high-$z$ AGN
were formed before $z \sim 4$. 

According to our model, it 
would be worth noting that quasar nuclei are hidden
by the dusty clouds unless the galactic wind could expel them from
the host galaxies. 
We also mention that any quasar nuclei are not necessarily to associate with
gas-rich circumnuclear environment though this 
implication is in contradiction to
what suggested for low-$z$ AGN (Yamada 1994).
Therefore, it seems very lucky that the CO  emission was
detected  from BR 1202$-$0725
at $z = 4.69$.

Finally, we revisit the important question: What is BR 1202$-$0725 ? 
As shown in section 3, the unambiguous CO detection from BR 1202$-$0725
is interpreted as an initial starburst galaxy which is forming  
either an elliptical
or a bulge with mass larger than $\sim 2 \times  10^{12} M_\odot$.
The elongated (Ohta et al. 1996) or the double-peaked (Omont et al. 1996)
CO distribution may be understood
as possible evidence for galactic wind in terms of our scenario.
If it is an elliptical galaxy, its formation epoch is estimated to be 
$z_{\rm f} \sim 5 {\rm -} 10$.
However, if it were a bulge former, 
the mass of bulge should be comparable with
that of typical ellipticals.
Since such massive bulges are rarer by two orders of magnitude than elliptical 
with similar masses (e.g., Woltier 1990),
the host of BR 1202$-$0725 may be an elliptical 
from a statistical ground.

\acknowledgments

We gratefully acknowledge T. Kodama and O. Nakamura 
for kindly providing us chemical evolution program packages.
Our special thanks to K. Ohta, T. Yamada, and R. McMahon for
fruitful discussions. We also thank T. Hasegawa and M. Honma 
for useful comments.
This work was financially supported in part by a Grant-in-Aid for
the Scientific Research by the Japanese Ministry of Education,
Culture, Sports and Science (Nos. 07044054 and 09640311).

\newpage

\begin{center}
{\bf Figure Captions}
\end{center}

\figcaption {
CO luminosity evolution for elliptical galaxies. The solid line 
represents the galactic wind model; $M_{\rm G}=2 \cdot 10^{12} M_{\odot}$,
$k=10$ Gyr$^{-1}$, $x=1.10$, $a=10$ Gyr$^{-1}$, $t_{\rm gw}=0.85$ Gyr,
and $z_f=10$. The dotted line gives a model with continuous star formation
and the dashed line shows a case for bound-wind model.
A filled square and a filled pentagon 
shows the observed $L_{\rm CO}$ of the high-$z$ quasar
BR 1202-0725 (Ohta et al. 1996) 
and the weak radio galaxy 53W002 (Scoville et al. 1997), respectively. 
Open triangles and open squares indicate the upper 
limits of negative detection from high-$z$ radio galaxies
by Evans et al. (1996) and van Ojik et al. (1997). 
Filled circles show the $L_{\rm CO}$ of 
nearby ellipticals (Wiklind \& Rydbeck 1986; Sage \& Wrobel 1989;
Wiklind \& Henkel 1989; Eckart et al. 1991; Lees et al. 1991; 
Sage \& Galletta 1993; Sofue \& Wakamatsu 1993; Wiklind, Combes \&
Henkel 1995) and open circles give the upper limits
for non-detected elliptical galaxies (Sofue \& Wakamatsu 1993; Wiklind, 
Combes, \& Henkel 1995).
\label {fig1}}

\figcaption {
CO luminosity evolution for spiral galaxies. The solid line
represents the bulge-disk model; $M_{\rm G}=2 \cdot 10^{11} M_{\odot}$,
$k=0.32$ Gyr$^{-1}$, $x=1.45$, $a=0.17$ Gyr$^{-1}$, and $z_f=10$ 
for the disc and $M_{\rm G}=10^{11} M_{\odot}$,
$k=10$ Gyr$^{-1}$, $x=1.10$, $a=10$ Gyr$^{-1}$, $t_{\rm gw}=0.36$ Gyr,
and $z_f=5$ for the bulge.
A filled square, a filled pentagon, open triangles, 
and open squares are the same as in Fig.1.
Small filled circles give the $L_{\rm CO}$
of nearby spiral galaxies taken from Braine et al. (1993).
\label{fig2}}

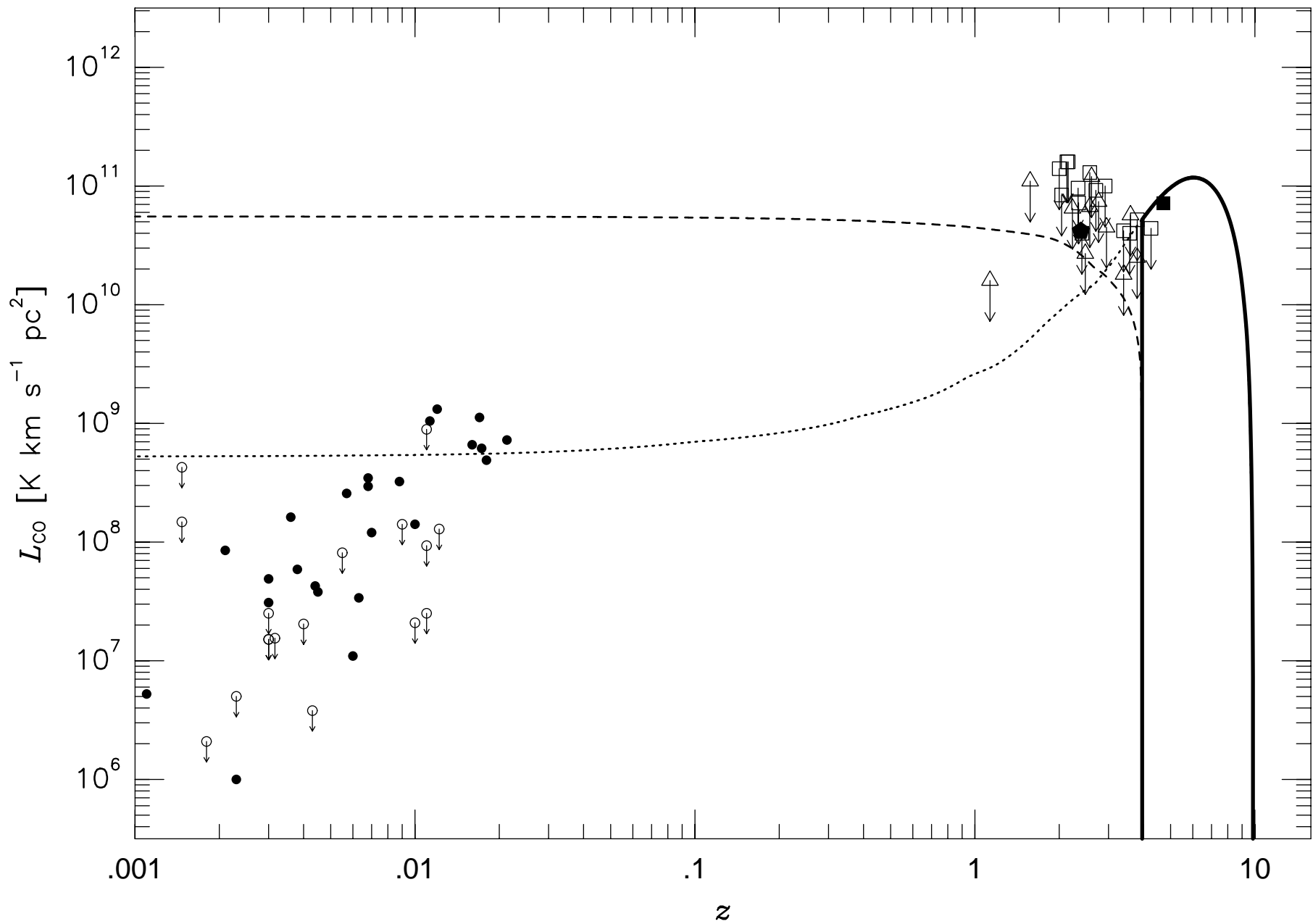
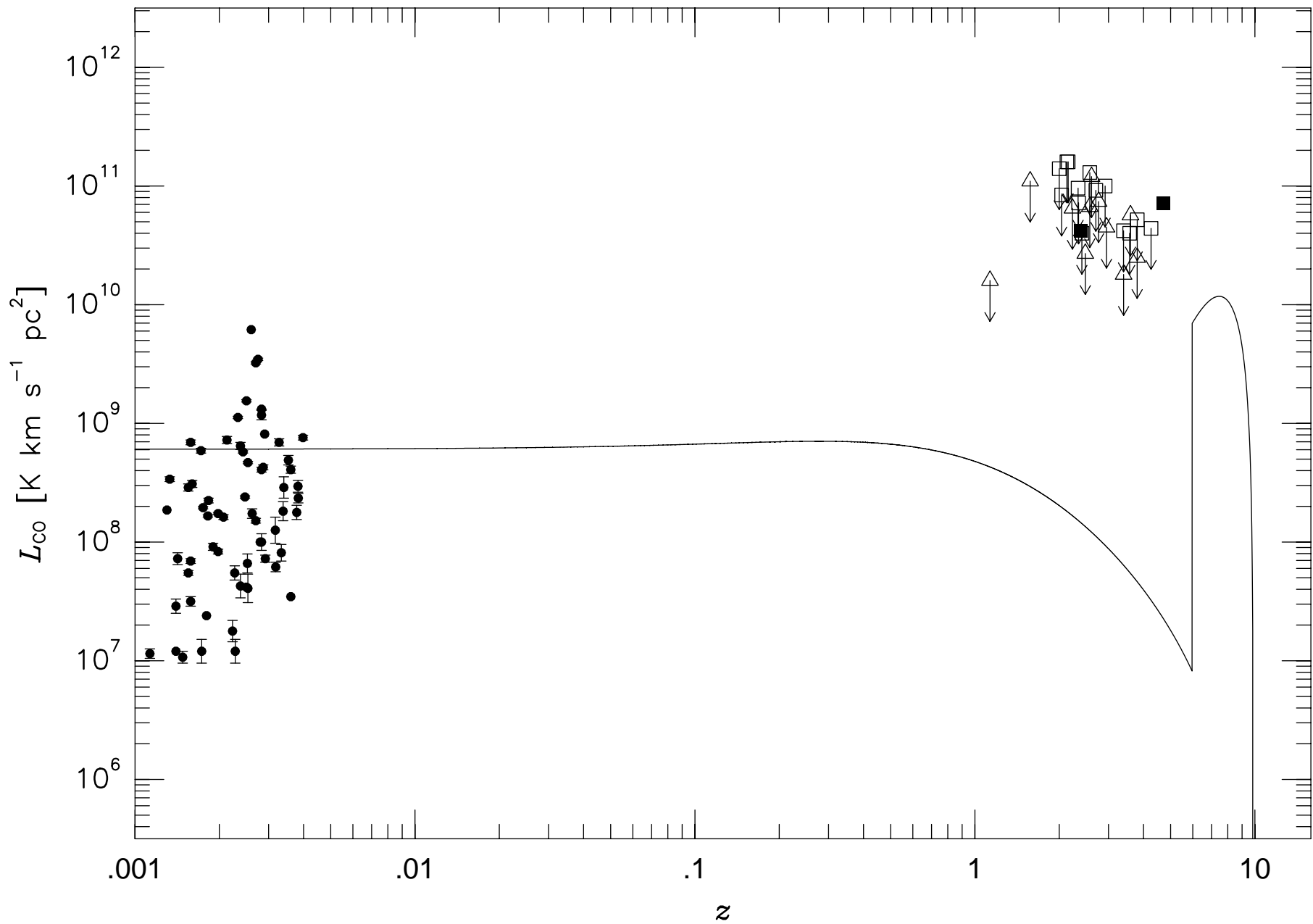
\end{document}